\documentclass[aps, pre, twocolumn, superscriptaddress]{revtex4-2}

\usepackage{graphicx}
\usepackage{amsmath}
\usepackage{amssymb}
\usepackage{amsfonts}
\usepackage{amsthm}
\usepackage{bm}
\usepackage{color}
\usepackage{xcolor}
\usepackage{soul}    
\usepackage{caption}
\usepackage{subcaption}
\usepackage{resizegather}
\usepackage[utf8]{inputenc}
\usepackage{url}
\usepackage{comment}

\begin{document}

\newcommand{\defeq}{\mathrel{\mathop:}=}

\title{Temperature fluctuations in finite systems:\\ Application to the one-dimensional Ising chain}

\author{Constanza Farías}
\email{m.fariasparra@uandresbello.edu}
\affiliation{Departamento de F\'isica, Facultad de Ciencias Exactas, Universidad Andres Bello. Sazi\'e 2212, piso 7, Santiago, 8370136, Chile.}

\author{Sergio Davis}
\affiliation{Research Center in the Intersection of Plasma Physics, Matter and Complexity (P$^2$mc), Comisión Chilena de Energía Nuclear, Casilla 188-D, Santiago, Chile}
\affiliation{Departamento de F\'isica, Facultad de Ciencias Exactas, Universidad Andres Bello. Sazi\'e 2212, piso 7, Santiago, 8370136, Chile.}

\date{\today}

\begin{abstract}
The theory of superstatistics, originally proposed for the study of complex nonequilibrium systems, has recently been extended to studies of small
systems interacting with a finite environment, because such systems display interestingly similar statistical behavior. In both situations there
are several applicable definitions of inverse temperature, either intrinsic or dependent of the statistical ensemble. In this work we develop these
concepts focusing our attention on a region of an isolated, one-dimensional Ising chain as an example of a subsystem that does not follow the canonical
Gibbs distribution. For this example, we explicitly show that superstatistics cannot describe the behavior of the subsystem, and verify a recently
reported relation between the fundamental and microcanonical inverse temperatures. Our results hint at a new framework for dealing with regions of
microcanonical systems with positive heat capacity, which should be described by some new class of statistical ensembles outside superstatistics
but still preserving the notion of temperature fluctuations.
\end{abstract}


\maketitle


\section{Introduction}

In traditional statistical mechanics we normally deal with small systems , such as particles interacting
inside an isolated box or otherwise in contact with a large
heat bath. Based on this, we either apply the microcanonical or canonical ensemble, respectively, to describe the system,
often making use of the fact that, in the thermodynamical limit, both ensembles are equivalent.

An interesting situation arises when we study a system in which there exists subsystem-environment interaction but only through nearest-neighbors, and where the size of
the environment is commensurate with the subsystem. In this case, these systems present temperature fluctuations, and several works\cite{Campi2005,Bustamante2005, Hill1998} treat these systems basically using the traditional statistical mechanics.

One of the most well known examples of small systems is the classical Ising model, in which nearest-neighbor interactions are given through
the exchange constant $J$. The classical one-dimensional Ising chain is described by the Hamiltonian
\begin{equation}
      \mathcal{H} = -J \sum_{i=1}^{N} s_{i}s_{i+1}. \label{eq:ising1D}
\end{equation}
where the spins $s_i \in \{-1, 1\}$ and we consider periodic boundary conditions.

Recently Ilin \emph{et al}~\cite{Ilin2020} showed how one can take a piece of $L$ spins from a microcanonical one-dimensional Ising chain of $N$ spins in total and describe it as a system in contact with an environment that is commensurate with the system. Their results show in detail how the system is described by a non-Gibbsian distribution that reduces to the
traditional canonical ensemble in the limit when $L \ll N$, as expected.

When dealing with systems presenting subsystem-environment interactions with a finite environment, a new theory known as \textit{superstatistics}~\cite{Beck2003, Beck2004} has been introduced. This theory was originally proposed to explain the statistics of different types of complex systems, such as plasmas and self-gravitating systems and, only lately, some applications to the thermodynamics of small systems\cite{Dixit2013, Dixit2015, Herron2021} characterized by short-range interactions have been presented.

In those cases, the system in contact with a finite environment has a statistical distribution of temperatures and does not follow the traditional Gibbs
distribution, i.e. the canonical ensemble. However, in the thermodynamical limit the uncertainty in temperature vanishes and the classical canonical ensemble is recovered.
Because this is exactly the situation in superstatistics, these authors proposed to use this framework, obtaining accurate fits.

However, it is not always the case that the framework of superstatistics is capable of describing these type of systems.
Accordingly, in this work we propose to use the properties of the fundamental and the microcanonical inverse temperatures to explore and study, as a particular case, the behavior
of a one-dimensional Ising subsystem, being a part of an isolated Ising chain. Under the constraints that the superstatistical theory imposes on the two temperature
functions, fundamental and microcanonical, we show that the non-Gibbsian distribution that describes the Ising subsystem is not consistent with superstatistics, despite having
temperature fluctuations. This opens up the idea of exploring the limits of the superstatistical framework and finding the family of statistical ensembles that describes this type of models. We show that these ensembles are characterized by a negative covariance between the microcanonical and fundamental temperatures, similar to previously reported results~\cite{Davis2021}, and hinting towards the existence of a new class of ensembles that are incompatible with
superstatistics.

%
%
%

The paper is organized as follows. In Section \ref{sec:superstat}, we briefly describe the theory of superstatistics and give the definition of the fundamental and microcanonical
inverse temperatures. In Section \ref{sec:results}, we give the theoretical results obtained by this analysis, presenting the statistical properties of the fundamental temperature, following by the application of the necessary condition that any superstatistical system must fulfill. Section~\ref{sec:discussion} provides a deeper discussion of
the results obtained, and finally in Section \ref{sec:conclusions}, we conclude this work.

\section{Superstatistics and its definition of temperature}
\label{sec:superstat}

Although the main objective of superstatistics~\cite{Beck2003, Beck2004} was originally to describe non-equilibrium systems in steady states, soon the framework was
extended~\cite{Dixit2013,Dixit2015,Davis2020}~in order to apply it to finite thermodynamical systems having energy and temperature fluctuations.

In the canonical ensemble, based on a fixed temperature $T$, and whose probability distribution of microstates is obtained by constraining the expectation value of energy according to the \textit{maximum entropy principle}~\cite{Jaynes1957,Jaynes2003}, we have,
\begin{equation}
    P(\bm{x}|\beta) = \frac{\exp(-\beta \mathcal{H}(\bm{x}))}{Z(\beta)}, \label{eq:densfunct}
\end{equation}
where $Z(\beta)=\int d\bm{x}\exp{(-\beta \mathcal{H}(\bm{x}))}$ is known as the \textit{partition function} and $\mathcal{H}(\bm{x})$ is the Hamiltonian of the system. On the other hand, for a non-equilibrium steady state system the probability density of microstates is of the form
\begin{equation}
    P(\bm{x}|S) = \rho(\mathcal{H}(\bm{x})), \label{eq:rho}
\end{equation}
where the function $\rho(E)$ is known as the \textit{generalized Boltzmann factor} or \emph{ensemble function}.

In superstatistics, which is a particular case of the form in Eq.~\ref{eq:rho}, a system experiences fluctuations of the inverse temperature
$\beta$, and therefore the probability distribution of microstates is replaced by a joint probability distribution using
\textit{Bayes' Theorem}~\cite{Jaynes2003},
\begin{equation}
    P(\bm{x},\beta) = P(\bm{x}|S)P(\beta|S) = \left[\frac{\exp(-\beta \mathcal{H}(\bm{x}))}{Z(\beta)}\right]P(\beta|S). \label{eq:jointprob}
\end{equation}

Eq.~\ref{eq:jointprob} shows an additional component beside the classical canonical ensemble, $P(\beta|S)$  known as the \textit{temperature distribution}, which
contains all the information of the temperature fluctuations into the system. In the particular case where
\begin{equation}
P(\beta|S)=\delta(\beta - \beta_{0}) \label{eq:canoncase}
\end{equation}
we recover the canonical ensemble, as it is expected.

Likewise, in principle we can consider the Hamiltonian of a composite system and extend the traditional formulation of nonequilibrium systems
by explicitly incorporating an environment $\bm{y}$ so that $\mathcal{H}(\bm x, \bm y) = H(\bm x)+G(\bm y)$, with a joint probability distribution given by
\begin{equation}
    P(\bm{x},\bm{y}|S)= \rho(H(\bm{x}) + G(\bm{y})) \label{eq:supercanonnew}
\end{equation}
in a steady state $S$, where $H(\bm{x})$ is the system Hamiltonian and $G(\bm{y})$ is the environment Hamiltonian.

In this work we will focus on the microcanonical ensemble, where for a system described by a Hamiltonian $\mathcal{H}(\bm{x})$ the energy is conserved, i.e. $\mathcal{H}(\bm{x}) = E_0$ for all relevant states $\bm x$. Because a given ensemble function $\rho(E)$ has a distribution of energies
\begin{equation}
P(E|S) = \rho(E)\Omega(E) \label{eq:probsuper}
\end{equation}
and we have  $P(E|E_0) = \delta(E-E_0)$, it follows that
\begin{equation}
    \rho(E) = \frac{\delta(E-E_{0})}{\Omega(E_0)}. \label{eq:microcanon}
\end{equation}

In the case of an isolated composite system, combining Eqs.~\ref{eq:supercanonnew} and \ref{eq:microcanon} yields
\begin{equation}
    P(\bm x, \bm y|E_0) = \frac{\delta(H(\bm x)+G(\bm y)-E_{0})}{\Omega(E_0)}. \label{eq:rhomicro}
\end{equation}

For every nonequilibrium steady state described by Eq.~\ref{eq:rho} we can obtain temperature by two different paths. First
we have an ensemble-dependent inverse temperature given by
\begin{equation}
\beta_F(E) \defeq -\frac{\partial}{\partial E}\ln \rho(E) \label{eq:defbF}
\end{equation}
known as the \textit{fundamental inverse temperature}. On the other hand, the intrinsic inverse temperature
\begin{equation}
\beta_\Omega(E) \defeq \frac{\partial}{\partial E}\ln \Omega(E) \label{eq:defbO}
\end{equation}
is the \textit{microcanonical inverse temperature}, related to the logarithm of the density of states, i.e. the Boltzmann entropy of the system. Being an
intrinsic temperature, that only depends on the definition of the Hamiltonian, it is in principle measurable in superstatistical systems and more general steady states. Both approaches to obtain the temperature become equivalent in the thermodynamical limit, however, their uncertainties, measured through their variances, are in general different.

\nocite{Velazquez2009}


\section{Results}
\label{sec:results}

\subsection{Fundamental temperature of a one-dimensional Ising subsystem}

In the following we present the calculation of the fundamental inverse temperature $\beta_F(E_s; E)$ for a subchain of $L$ Ising spins, which is part of an
isolated Ising chain of $N$ spins at total energy $E$. First, we take Eq. 6 from Ref.~\cite{Ilin2020}, which gives
the number of states $\omega_n$ having a fixed subsystem energy,
	\begin{equation}
		\omega_{n} = \frac{1}{\Omega_M}\frac{(N-L+1)!}{(M-K)!(N-L+1-M+K)!} \label{eq:ilinprobdens}
	\end{equation}

where $\Omega_M$ is the number of states with total energy compatible with $M$, and the entire chain has fixed energy $E$. Here, the integer variables $K$ and $M$ determine the subsystem and system energy, respectively, through the relations
\begin{align}
K \defeq & \;\frac{L}{2} - \frac{1}{2}+ \frac{E_s}{2J}, \label{eq:K}\\
M \defeq & \;\frac{N}{2} + \frac{E}{2J}. \label{eq:M}
\end{align}
and in our case we will take $\rho(E_s; E)$ as the limit of $\omega_n$ when $N \rightarrow \infty$. In addition, we will
consider the choice $J = 1$ to simplify the computation in the rest of this work. We will now define the intensive quantities
\begin{equation}
\gamma := \frac{L}{N}
\end{equation}
and
\begin{equation}
\epsilon_s \defeq \frac{E_s}{N}, \qquad \epsilon \defeq \frac{E}{N},
\end{equation}
so that we can take the thermodynamic limit ($N\rightarrow \infty$) while preserving the relative proportions of subsystem and environment.

As it is shown in detail in the appendix, using the property $\psi(n) = H_{n-1}-\gamma_e$ where $\psi(n)$ is the digamma function, and the asymptotic expansion of the
harmonic number $H_n$\cite{Riley2002},
\begin{equation}
H_n \sim \ln(n) + \frac{1}{2n} + \gamma_e
\end{equation}
where $\gamma_e \defeq$ 0.57721566\ldots ~is known as the \textit{Euler-Mascheroni} constant, we finally get
%
%
\begin{equation}
		\beta_{F} = \frac{1}{2}\ln \left(\frac{\displaystyle{1+(\epsilon_{s}-\epsilon) - \gamma}}{\displaystyle{1+(\epsilon-\epsilon_{s}) - \gamma}}\right).
		\label{eq:bF}
\end{equation}

We see that the fundamental inverse temperature is an intensive quantity, as expected. Furthermore, when applying the thermodynamic limit together with $\gamma \longrightarrow 0$, we have $\epsilon_s \ll \epsilon$ (because $L \ll N$) and we recover the Gibbs distribution, with the same inverse temperature
as in Ref.~\cite{Ilin2020},
	\begin{equation}
		\beta = \frac{1}{2} \ln \left(\frac{1-\alpha}{\alpha}\right), \label{eq:Gibbsb}
	\end{equation}
where $\alpha \defeq \frac{1}{2}(1+ E/NJ)$. 
In the case when $\gamma \longrightarrow 1$, the fundamental inverse temperature $\beta_F$ is indeterminate as expected, this is because the sub-system has the same size as the entire system, therefore it is an isolated system with ensemble function proportional to a Dirac delta function~\cite{Velazquez2009}.


\subsection{Connection between the fundamental and microcanonical temperatures}

As follows from Eq.~\ref{eq:rhomicro} for a subsystem of a microcanonical ensemble, it corresponds an ensemble function
\begin{equation}
\begin{split}
\rho^{(\bm x)}(E; E_0) & = \frac{1}{\Omega(E_0)}\int d\bm{y} \delta(E + G(\bm y)-E_0) \\
                       & = \frac{\Omega^{(\bm y)}(E_0-E)}{\Omega(E_0)},
\end{split}
\end{equation}
and from this, by taking the logarithmic derivative of $\rho^{(\bm x)}(E; E_0)$ with respect to $E$ we can obtain the fundamental inverse temperature of the subsystem as
\begin{equation}
\begin{split}
    \beta_{F}^{(\bm{x})}(E; E_0) & = -\frac{\partial}{\partial E}\ln \rho^{(\bm x)}(E; E_0) \\
                                 & = \beta_{\Omega}^{(\bm{y})}(E_{0}-E) \label{eq:bfbo}
\end{split}
\end{equation}
where $\bm{x}$ refers to the microstates of the system and $\bm{y}$ to the environment microstates. We clearly see that there exists a relation between these two inverse temperatures, the fundamental temperature of the subsystem and the microcanonical temperature of the environment,

In our case, by explicitly computing $\beta_\Omega$ from the density of states of the Ising model, we obtained
\begin{equation}
    \beta_{\Omega} =\frac{1}{2}\ln \left(\frac{\displaystyle{\gamma - \epsilon_{s}}}{\displaystyle{\gamma + \epsilon_{s}}}\right) \label{eq:betaomega}
\end{equation}
by which we can verify the relation in
Eq.~\ref{eq:bfbo}, simply by performing the replacements
\begin{subequations}
\begin{align}
\gamma & \longrightarrow (1-\gamma), \\
\epsilon_{s} & \longrightarrow (\epsilon-\epsilon_{s}),
\end{align}
\end{subequations}
into Eq.~\ref{eq:betaomega}, after which we recover Eq.~\ref{eq:bF}, the fundamental inverse temperature.

We see that this transformation is completely general and can be used in several ways according to the choice of subsystem and environment, noting that, if $\gamma$ is
the relative size of the subsystem, then $1-\gamma$ corresponds to the relative size of the environment. Moreover, if $E_s$ is the energy of the subsystem, then $E-E_s$
is the energy of the environment when $E$ is the fixed total energy.



Due to the fact that, for every superstatistical model it must hold that
\begin{equation}
\frac{\partial \beta_F(E)}{\partial E} = -\big<(\delta \beta)^2\big>_{E,S} \leq 0,
\end{equation}
it follows from Eq.~\ref{eq:bfbo} that any system who is in contact with an environment with $C_{V} > 0$ will not follow superstatistics ~\cite{Davis2021}, because we have
\begin{equation}
	\frac{\partial \beta_{F}^{(\bm{x})}}{\partial E} = \frac{\big(\beta_{\Omega}^{(\bm{y})}\big)^{2}}{C_{V}^{(\bm{y})}} \label{eq:heatcap}
\end{equation}
where we have used
\begin{equation}
C_V(\varepsilon) = \left(\frac{\partial T(\varepsilon)}{\partial \varepsilon}\right)^{-1} =  -\frac{\beta_{\Omega}(\varepsilon)^2}{\beta'_{\Omega}(\varepsilon)}.
\end{equation}

\noindent
In our case, we have
%
%
\begin{equation}
\label{eq:partialbF}
	\frac{\partial \beta_{F}}{\partial \epsilon_s} = \frac{1}{2}\left[\frac{1}{1+(\epsilon-\epsilon_s) - \gamma} + \frac{1}{1+(\epsilon_s-\epsilon) - \gamma}\right], 
\end{equation}
which we can show is always positive, therefore the model does not follow superstatistics. This is because, in order for $\beta_F$ in Eq.~\ref{eq:bF} to be
a real number, we need both denominators to be of equal sign, and in fact they are both positive, as we show below. From Eq.~\ref{eq:K} and 
Eq.~\ref{eq:M} we can express the denominators in Eq.~\ref{eq:partialbF} in terms of the original variables $M$ and $K$, and taking into account the arguments of 
the factorials in the denominator of $\omega_n$ in Eq.~\ref{eq:ilinprobdens}, we have the inequalities
\begin{align}
M-K > & \; 0, \label{eq:ineq1}\\
1-L-M+N+K > & \; 0. \label{eq:ineq2}
\end{align}

\noindent
From Eq.~\ref{eq:ineq1} we immediately obtain 
\begin{equation}
\frac{M-K}{N} = \frac{1}{2}(1-\gamma + \epsilon-\epsilon_s) > 0 \label{eq:defMK}
\end{equation}
while from Eq.~\ref{eq:ineq2} it follows the inequality
\begin{equation}
M-K < \; N-L+1 \approx N-L, \label{eq:ineq3}
\end{equation}
where we have approximated $N+1 \approx N$ because of thermodynamic limit. Hence, we have
\begin{equation}
\frac{M-K}{N} = \frac{1}{2}(1-\gamma+\epsilon-\epsilon_s) < 1-\gamma, \label{eq:ineq4}
\end{equation}
so that
\begin{equation}
    (1-\gamma)+(\epsilon-\epsilon_s) < 2(1-\gamma) \label{eq:absgamma}
\end{equation}
and then it follows that
\begin{equation}
\epsilon-\epsilon_s < 1-\gamma. \label{eq:absgamma2}
\end{equation}

\noindent
Eqs.~\ref{eq:defMK} and~\ref{eq:absgamma2} prove that the right-hand side of Eq.~\ref{eq:partialbF} is always positive.

As we have mentioned, superstatistics for a region of an isolated system requires the environment to have negative heat capacity, which explains the result just obtained.

This negative heat capacity can be shown to ocurr in both systems with long-range interactions~\cite{Thirring1970, Campa2009},
as well as in short-range interactions where system size and energies are comparable with an environment\cite{Umirzakov1999, Chomaz2002, Eryurek2007, Eryurek2008}.
In both situations the key thermodynamical feature is the presence of a region of convex entropy~\cite{Latella2015, Latella2017}.

In the case of microcanonical finite systems this behavior is clear in first-order phase transitions which are characterized by a bimodal energy distribution and anomalously large fluctuations.
In fact, a bimodal energy distribution, originating from particular features of the energy landscape, is a necessary and sufficient condition for a system to show 
negative heat capacity~\cite{Schmidt2000,Carignano2010}, and in the case of classical spin systems, the Potts model provides some examples of this kind of 
behavior~\cite{Moreno2018,Farias2021}.

\subsection{Variance and correlations of the fundamental temperature}

Given that we have just shown the Ising subsystem does not follow superstatistics, this does not means that it follows the traditional Gibbs distribution, instead, it for sure follows another type of statistics, because $\beta_F$ in Eq.~\ref{eq:bF} is not the constant function. In the following, we will show that there exist a possibility to constrain the kind of distribution, providing additional bounds in the theory of superstatistics.

Consider the inverse temperature covariance parameter $\mathcal{U}$, defined as
%
\begin{equation}
\mathcal{U} \defeq \Big<\delta \beta_F\delta \beta_\Omega\Big>_S.
\end{equation}

\noindent In the limit of small, Gaussian fluctuations of energy, that is, when
\begin{equation}
P(E|S) \approx \frac{1}{\sqrt{2\pi}\sigma_E}\exp\Big(-\frac{(E-E^*)^2}{2\sigma_E^2}\Big) \label{eq:gaussian}
\end{equation}
with $\sigma_E^2 \defeq \big<(\delta E)^2\big>_S$, we can approximate
\begin{equation}
\big<\delta f\,\delta g\big>_S \approx f'(E^*)g'(E^*)\big<(\delta E)^2\big>_S   \label{eq:fg}
\end{equation}
for any pair of functions $f$ and $g$ of the energy \cite[p.~51]{BailerJones2017}. Hence we will have
\begin{equation}
    \mathcal{U} \approx \beta'_{F}(E^{*})\beta'_{\Omega}(E^{*})\left<(\delta E)^{2}\right>.
\end{equation}
Note that $E^{*}$ is the most probable energy of the system and is given by the equality
\begin{equation}
\beta_F(E^*) = \beta_\Omega(E^*),
\end{equation}
which leads to the intuitive result,
\begin{equation}
\epsilon_{s}^{*} = \gamma \epsilon.
\end{equation}

Moreover, for any steady state ensemble following Eq.~\ref{eq:probsuper}, in the approximation of Eq.~\ref{eq:gaussian}, the derivatives of the
microcanonical and fundamental inverse temperature at $E^*$ are connected to the variance of the energy by
\begin{equation}
    \left<(\delta E)^{2}\right> = \frac{1}{\beta'_{F}(E^*) - \beta'_{\Omega}(E^*)}. \label{eq:varE}
\end{equation}

\noindent
We can obtain the variance of both the microcanonical inverse temperature $\sigma_{\beta_{\Omega}}^{2}$ and the fundamental inverse temperature
$\sigma_{\beta_{F}}^{2}$ by using Eq.~\ref{eq:fg} for $f = g$, 
\begin{subequations}
\begin{eqnarray}
    \label{varbo}
    \sigma_{\beta_{\Omega}}^{2} & =  \left<(\delta E)^{2}\right> |\beta'_{\Omega}(\epsilon^{*})|^{2} & = {\displaystyle \frac{1}{|N\gamma (\epsilon^{2}-1)|}}, \\
    \label{varbf}
    \sigma_{\beta_{F}}^{2} & =  \left<(\delta E)^{2}\right> |\beta'_{F}(\epsilon^{*})|^{2} & = {\displaystyle \frac{1}{|N(\gamma-1)(\epsilon^{2}-1)|}}
\end{eqnarray}
\end{subequations}
respectively, where we have used Eq.~\ref{eq:partialbF} and the derivative of Eq.~\ref{eq:betaomega} at $\epsilon^* = \gamma\epsilon$,
\begin{equation}
    \frac{\partial \beta_{\Omega}}{\partial \epsilon_{s}} = \frac{\gamma}{(\epsilon_{s}^{*2} - \gamma^{2})} = \frac{1}{\epsilon^2-1}. \label{eq:14}
\end{equation}

\noindent
Replacing our previous results, we obtain in our case that Eq.~(\ref{eq:varE}) reduces to
\begin{equation}
     \left<(\delta E)^{2}\right> = N\gamma(\gamma - 1)(\epsilon^{2}-1). \label{eq:evar}
\end{equation}

\noindent
Finally, in this particular model, we obtain
\begin{equation}
    \mathcal{U} \approx \frac{1}{N}\left[\frac{1}{(\epsilon^{2}-1)}\right] < 0  \label{eq:finalU}
\end{equation}
for large $N$.
\section{Discussion}
\label{sec:discussion}

Noting the fact that the variance of energy in Eq.~(\ref{eq:evar}) is proportional to $N$, unlike the inverse temperature derivatives,
which are each proportional as $1/N$, we can confirm the asymptotic dependence
\[\mathcal{U} \propto \frac{1}{N}\] that is expected of the variance of an intensive quantity. Additionally, here we note that, because $\sigma_{E}^{2} > 0$ and $0 \leq \gamma \leq 1$, that is, $\gamma - 1 < 0$, it follows from Eq.~\ref{eq:evar} that is strictly necessary that $\epsilon < J$ must be fulfilled, which is in fact true of the Ising
chain but was not used in the analysis.

The result given by Eq.~\ref{eq:finalU} is an interesting one, because it can be clearly seen that $\mathcal{U}$ is independent of the $\gamma$ parameter, that is, is independent of the portion of the entire system that we are considering as the subsystem, i.e. it follows that
\begin{equation*}
    \mathcal{U}_{\text{sub}} = \mathcal{U}_{\text{env}} = \mathcal{U}_{\text{sys}}.
\end{equation*}

\begin{figure}[h!]
 \centering
 \includegraphics[scale=0.55]{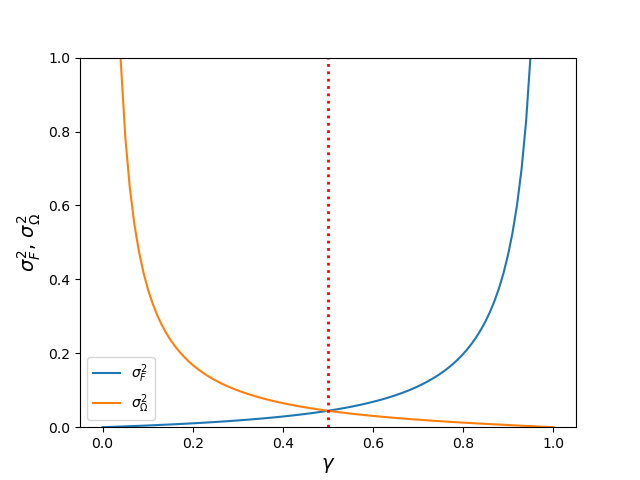}
 \hfill
    \caption{Relation between the variance of each inverse temperature with the $\gamma$ parameter and $\epsilon = 0.5$. There exists a crossover at $\gamma = \gamma_c = 0.5$, where both variances are equal.}
  \label{fig_n1}
\end{figure}

In Fig.~\ref{fig_n1} we can see the behavior of the variances of the microcanonical and fundamental inverse temperatures, given by Eqs.~\ref{varbo} and ~\ref{varbf}, 
respectively. It is clear from our results that there is a crossover of the curves exactly at the value  \[\gamma_c \defeq \frac{1}{2},\] so that when 
$\gamma < \gamma_c$ we have $\sigma_{F}^{2} < \sigma_{\Omega}^{2}$, which suggests a kind of conjugate relationship between both $\beta_F$ and $\beta_\Omega$, where they cannot 
be simultaneously determined with precision. This is linked to the exchange symmetry between the relative sizes of the subsystem and the environment, namely 
$\gamma \rightarrow 1-\gamma$, in such a way that, when both regions of the system are equal in size, at $\gamma = 1/2$, the variances are equal as expected.



%

\begin{figure}[h!]
 \centering
 \includegraphics[scale=0.50]{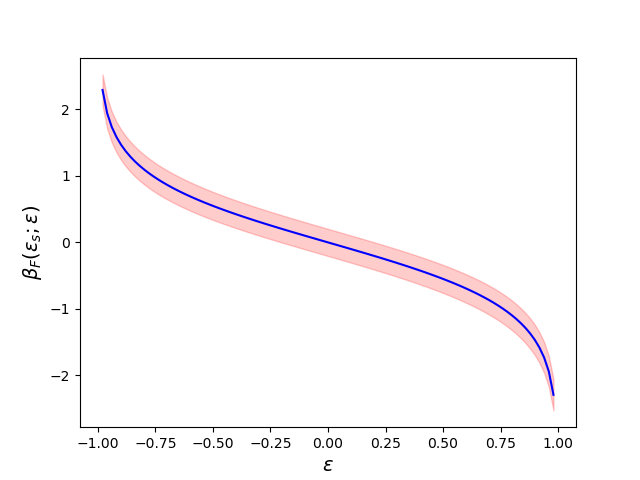}
 \hfill
\caption{Fundamental inverse temperature of an Ising subsystem (blue line) as a function of the total energy of the system at $\gamma = 0.5$. The light-red band indicates the uncertainty of $\beta_F$ according to Eq.~\ref{varbf}.}
 \label{fig_n2}
\end{figure}

\vspace{5pt}
According to Fig.~\ref{fig_n2}, for any isolated system it is, in principle, simple to compute the microcanonical caloric curve just by considering the definition
of $\beta'_{\Omega}$, this agrees with the analytical solution of the one-dimensional Ising model in the canonical ensemble~\cite[p.~122]{Chandler1987},
where the partition function is given by
\begin{equation}
   Z(\beta) = \left[2\cosh{(\beta J)}\right]^{N}
\end{equation}
and the corresponding caloric curve is
\begin{equation}
    E(\beta) \defeq -\frac{\partial}{\partial \beta}\ln{Z(\beta)} = -JN\tanh{\beta J}.
\end{equation}

\noindent
By inverting $E(\beta)$ and applying the relation
\begin{equation*}
    \text{atanh}(x) = \frac{1}{2}\ln\left({\frac{1+x}{1-x}}\right)
\end{equation*}
we readily obtain
\begin{equation}
\beta(E) = \frac{1}{2J}\ln \left(\frac{1-\epsilon/J}{1+\epsilon/J}\right),
\end{equation}
which is the same result obtained by replacing $\gamma=1$, $\epsilon_s=\epsilon$ and $J=1$ in Eq.~\ref{eq:betaomega}.

Fig.~\ref{fig_n2} shows the fundamental inverse temperature $\beta_F$ in Eq.~\ref{eq:bF} and its uncertainty as a function of $\epsilon$ in the entire
range $\epsilon \in [-1, 1]$. We obtained an inflection point at $\epsilon = 0$, where the curvature changes. We can also see a symmetry between the fundamental inverse temperature $\beta_{F}$ and the energy of the system $\epsilon$, namely
\begin{equation}
\beta_F(-\gamma \epsilon; -\epsilon) = -\beta_F(\gamma \epsilon; \epsilon),
\end{equation}
which implies $\beta_F = 0$ at $\epsilon = 0$ for any $\gamma$.

Then, for positive values of the total energy, the fundamental inverse temperature become negative, as expected
for a system with an upper bound of energy, with the same uncertainty as in the positive branch. For values of $\varepsilon$ such that $|\epsilon| \rightarrow 1$, the uncertainty tends to zero, as expected.



\section{Concluding Remarks}
\label{sec:conclusions}

In this work we have used the finite, classical, one-dimensional Ising model to explore the temperature fluctuations in a system in contact with a commensurate environment, described by
a non-Gibssian distribution, as reported in Ref.~\cite{Ilin2020}. We proved, by computing the fundamental inverse temperature and its derivative, that the Ising subchain of $L$ spins inside a microcanonical chain of $N$ spins cannot be described within the superstatistical framework. According to this last result, we focus our analysis on the behavior of the fundamental and microcanonical inverse temperatures, where we obtained a dependence on the energy $E_s$ and the relative size $\gamma = L/N$.

We have also highlighted the importance of an environment with negative heat capacity, expected in systems with long-range interactions, as a necessary condition for superstatistics, providing a new, simpler method to quickly recognize if a particular setup can be described by superstatistics. 

From our results we see that there is an invariance of the inverse temperature covariance $\mathcal{U}$ between the full system, the subsystem and the environment, that is, its value is independent of the portion of the system we take. This suggests the possibility of describing some aspects of the entire system using information from an arbitrary subsystem, possibly leading to the use of another family of statistical models, different from the ones employed in traditional statistical mechanics and from some well-known generalizations of statistical mechanics such as Tsallis statistics.

By using a Gaussian approximation in the limit of small fluctuations, we computed the uncertainty of the fundamental and microcanonical inverse temperatures which have a conjugate behavior given by the symmetry relation between the sub-system and environment.

In summary, this work presents an alternative method for the generalized thermodynamics of small systems in contact with a commensurate environment, strongly suggesting that when a system experiments temperature and energy fluctuations a description using superstatistics is not always possible.

\section{Acknowledgements}

The authors gratefully acknowledge funding from ANID FONDECYT Regular 1220651. CF also acknowledges Beca ANID Doctorado Nacional/(2021) - 21210658.


\newpage
\appendix
\title{Appendix A}
\appendix

\section{Derivation of the fundamental temperature of the Ising chain}

In order to show the step-by-step computation of how the fundamental inverse temperature $\beta_F$ was obtained, consider the following definitions
which relate the original parameters with the energy of the system $E$ and the sub-system $E_s$, that is,
\begin{equation}
\frac{M}{N} = \frac{1}{2}+\frac{E}{2N},
\end{equation}
\begin{equation}
K = \frac{1}{2}\left(\gamma N - 1 + E_s\right),
\end{equation}
and $L=\gamma N$. We start with the quantity $\omega_n$ of Ref.~\cite{Ilin2020}, which in our case corresponds to the ensemble function $\rho(E_s; E)$


{\small
	\begin{align}
		\rho(E_s) = & \frac{\Gamma\Big(N(1-\gamma)+2\Big)}{\Gamma\Big(\frac{3}{2}+\frac{1}{2}\left(E-E_s\right)+\frac{N}{2}\left(1-\gamma)\right)\Big)} \times \nonumber 
		\\&
		\frac{1}{\Gamma\Big(\frac{3}{2}+\frac{1}{2}(E_s - E)+\frac{N}{2}(1-\gamma)\Big)}	\label{eq:A3}
		\end{align}
}


The next step was to take the logarithmic derivative of $\rho$ in Eq.~(\ref{eq:A3}), with respect to the sub-system energy,
%
%

	\begin{align}
		\frac{\partial\ln \rho(E_s)}{\partial E_s} = & \frac{1}{2}\Bigg[\psi\left(\frac{1}{2}\big(3+E-E_s+ N(1-\gamma)\big)\right) - \nonumber
		\\ &
		\psi\left(\frac{1}{2}\big(3-E+E_s+ N(1-\gamma)\big)\right)\Bigg]
    \label{eq:A4}
	\end{align}

Applying the definition of $\beta_F$ and simplifying, we obtained the last expression in terms of the
harmonic number $H_n$, finally we obtained a simplified expression for $\beta_{F}$
%

%
	\begin{align}
		\beta_{F} = &  \frac{1}{2}\ln{\Bigg[\frac{N(1-\gamma)+(E_s-E)}{N(1-\gamma)+(E-E_s)}\Bigg]},  \label{eq:A6}
	\end{align}
%
that, when cancelling $N$ and using $\epsilon_s \defeq E_s/N$, $\epsilon \defeq E/N$ reduces to Eq.~\ref{eq:bF}.








\bibliographystyle{unsrt}
\bibliography{references}

\begin{thebibliography}{10}

\bibitem{Campi2005}
X.~Campi and H.~Krivine.
\newblock Partial energy fluctuations and negative heat capacities.
\newblock {\em Phys. Rev C}, 71:041601(R), 2005.

\bibitem{Bustamante2005}
C.~Bustamante, J.~Liphardt, and F.~Ritort.
\newblock The nonequilibrium thermodynamics of small systems.
\newblock {\em Physics Today}, 58:43, 2005.

\bibitem{Hill1998}
T.~L. Hill and R.~V. Chamberlin.
\newblock Extension of the thermodynamics of small systems to open metastable
  states: An example.
\newblock {\em Proc. Natl. Acad. Sci.}, 95:12779--12782, 1998.

\bibitem{Ilin2020}
P.~K. Ilin, G.~V. Koval, and A.~M. Savchenko.
\newblock A non-{G}ibbs distribution in the {I}sing model.
\newblock {\em Theoretical and Mathematical Physics}, pages 415--419, 2020.

\bibitem{Beck2003}
C.~Beck and E.G.D. Cohen.
\newblock Superstatistics.
\newblock {\em Phys. A}, 322:267--275, 2003.

\bibitem{Beck2004}
C.~Beck.
\newblock Superstatistics: theory and applications.
\newblock {\em Cont. Mech. Thermodyn.}, 16:293--304, 2004.

\bibitem{Dixit2013}
P.D. Dixit.
\newblock A maximum entropy thermodynamics of small systems.
\newblock {\em Phys. Chem.}, 138:184111, 2013.

\bibitem{Dixit2015}
P.D. Dixit.
\newblock Detecting temperature fluctuations at equillibrium.
\newblock {\em Phys. Chem.}, 17:13000--13005, 2015.

\bibitem{Herron2021}
L.~Herron and P.D. Dixit.
\newblock Thermal statistics of small magnets.
\newblock {\em Journal of Statistical Mechanics: Theory and Experiment},
  2021(3):033207, 2021.

\bibitem{Davis2021}
S.~Davis.
\newblock Fluctuating temperatures outside superstatistics: Thermodynamics of
  small systems.
\newblock {\em Phys. A}, 589:126665, 2021.

\bibitem{Davis2020}
S.~Davis.
\newblock On the possible distributions of temperature in nonequilibrium steady
  states.
\newblock {\em J. Phys. A: Math. Theor.}, 53:045004, 2020.

\bibitem{Jaynes1957}
E.~T. Jaynes.
\newblock Information theory and statistical mechanics.
\newblock {\em Phys. Rev.}, 106:620, 1957.

\bibitem{Jaynes2003}
E.~T. Jaynes.
\newblock {\em "Probability theory: The Logic Of Science"}.
\newblock Cambridge University Press, 2003.

\bibitem{Velazquez2009}
L.~Velazquez and S.~Curilef.
\newblock A thermodynamic fluctuation relation for temperature and energy.
\newblock {\em J. Phys. A: Math. Theor.}, 42:095006, 2009.

\bibitem{Riley2002}
K.~F. Riley, M.~P. Hobson, and S.~J. Bence.
\newblock {\em Mathematical Methods for Physics and Engineering: A
  Comprehensive Guide}.
\newblock Cambridge University Press, 2002.

\bibitem{Thirring1970}
W.~Thirring.
\newblock Systems with negative specific heat.
\newblock {\em Z. Physik}, 235:339--352, 1970.

\bibitem{Campa2009}
A.~Campa, T.~Dauxois, and S.~Ruffo.
\newblock Statistical mechanics and dynamics of sovable models with long-range
  interactions.
\newblock {\em Phys. Rep.}, 480:57--159, 2009.

\bibitem{Umirzakov1999}
I.~H. Umirzakov.
\newblock van der {W}aals type loop in microcanonical caloric curves of finite
  systems.
\newblock {\em Physical Review E}, 60:7550--7553, 1999.

\bibitem{Chomaz2002}
P.~Chomaz and F.~Gulminelli.
\newblock {\em Phase Transitions in Finite Systems}.
\newblock Springer Berlin Heidelberg, 2002.

\bibitem{Eryurek2007}
M.~Eryürek and M.H. Güven.
\newblock Negative heat capacity of {A}r$_{55}$ cluster.
\newblock {\em Physica A}, 377:514--522, 2007.

\bibitem{Eryurek2008}
M.~Eryürek and M.~H. Güven.
\newblock Peculiar thermodynamic properties of {L}{J}$_n$ ($n$=39-55) clusters.
\newblock {\em European Physical Journal D}, 48:221--228, 2008.

\bibitem{Latella2015}
I.~Latella, A.~Perez-Madrid, A.~Campa, L.~Cassetti, and S.~Ruffo.
\newblock Thermodynamics of nonadditive systems.
\newblock {\em Phys. Rev. Letters}, 114:230601, 2015.

\bibitem{Latella2017}
I.~Latella, A.~Perez-Madrid, A.~Campa, L.~Cassetti, and S.~Ruffo.
\newblock Long-range interacting systems in the unconstrained ensemble.
\newblock {\em Phys. Rev. E.}, 95:012140, 2017.

\bibitem{Schmidt2000}
M.~Schmidt, R.~Kusche, T.~Hippler, J.~Donges, W.~Kronmüler, B.~von Issendorff,
  and H.~Haberland.
\newblock Negative heat capacity for a cluster of 147 sodium atoms.
\newblock {\em Phys. Rev. Letters}, 86:1191, 2000.

\bibitem{Carignano2010}
M.~A. Carignano and I.~Gladich.
\newblock Negative heat capacity of small systems in the microcanonical
  ensemble.
\newblock {\em EPL}, 90:63001, 2010.

\bibitem{Moreno2018}
F.~Moreno, S.~Davis, C.~Loyola, and J.~Peralta.
\newblock Ordered metastable states in the {P}otts model and their connection
  with the superheated solid state.
\newblock {\em Phys. A}, 509:361--368, 2018.

\bibitem{Farias2021}
C.~Farías and S.~Davis.
\newblock Multiple metastable-states in an off-lattice {P}otts model.
\newblock {\em Phys. A}, 581:126215, 2021.

\bibitem{BailerJones2017}
Coryn~A.L. Bailer-Jones.
\newblock {\em Practical Bayesian Inference: A primer for Physical Scientists}.
\newblock Cambridge University Press, Oxford, UK, 2017.

\bibitem{Chandler1987}
D.~Chandler.
\newblock {\em Introduction to modern Statistical Mechanics}.
\newblock Oxford University Press, 1987.

\end{thebibliography}

\end{document}